# A liquid crystal geometric phase hyperbolic lens with a positive focal length for arbitrary circularly polarized incidence

Boyuan Li,[1] Xiaoqian Wang,[1,3] Kean Zhu,[2] Dong Shen,[1] Zhigang Zheng[1,*]

[1]School of Physics, East China University of Science and Technology, Shanghai 200237, China
[2]Electronic Engineering, East China University of Science and Technology, Shanghai 200237, China
[3]xqwang@ecust.edu.cn
*zgzheng@ecust.edu.cn

ABSTRACT

This paper presents the design and experimental validation of a liquid crystal geometric phase hyperbolic lens (LCHL) with positive focal lengths for arbitrary circularly polarized light. Utilizing Pancharatnam-Berry phase modulation, the lens enables isotropic focusing for both left- and right-handed circular polarization. We fabricated and characterized two lenses with different focal lengths, and their diffraction patterns were analyzed experimentally. The results exhibit strong agreement with theoretical simulations, highlighting the lens's capability for precise optical field modulation. The proposed LCHL demonstrates significant potential for detecting weak polarization states, making it a promising tool for advanced applications in biomedical imaging, optical vortex generation, and multifocal optical systems.

The origins of lenses can be traced back to ancient civilizations. In early Egypt and Mesopotamia, people may have accidentally discovered natural crystals and glass artifacts, which were then used as simple magnification tools. Greek philosophers such as Aristotle and Euclid studied the reflection and refraction of light, laying the foundation for optical theory. By the time of ancient Rome, the magnifying function of lenses was already known, as evidenced by the Roman philosopher Descartes's description of using a bowl of water or a glass sphere to magnify objects[1,2]. However, the true development of lenses began in the late Middle Ages when Italian craftsmen started manufacturing eyeglasses. This technology advanced further during the Renaissance, with lenses becoming central components in optical instruments such as telescopes and microscopes.

With technological advancements, optical lenses have become increasingly prevalent in both daily life and scientific research, owing to their diverse shapes and corresponding characteristics and applications. For instance, spherical lenses, which can have convex, concave, or planar surfaces, not only function as collimating lenses in fiber optic collimators[3] but are also widely used in eyeglasses, microscopes, and other everyday items. Cylindrical lenses, characterized by their ability to focus or diverge light along a single axis (usually vertical or horizontal) while leaving light in the other direction unaffected, are commonly used in laser processing and beam shaping[4,4]. Additionally, axicons, which can focus light into a line or a ring, are often employed to convert incoming Gaussian beams into non-diffracting Bessel beams[5,6]. Bessel beams, known for their self-healing and non-diffracting properties, maintain their shape and structure even over long distances[7], making axicons valuable for application in telescopes, autocollimators, microscopes[8], and laser surgery[6].

Beyond these types of lenses, there are also Fresnel lenses and diffraction gratings. Fresnel lenses, characterized by their concentric rings or serrated surfaces, are used for focusing or dispersing light[9] and are commonly applied in solar lighting and projection systems[10]. Diffraction gratings, on the other hand, employ micro-optical structures to diffract light, achieving focusing or beam shaping without relying on refraction. These optical lenses, whether used individually or in combination, serve various purposes, such as focusing light onto sensors in camera lenses or correcting vision problems like myopia, hyperopia, astigmatism, or presbyopia in eyeglasses. They are also integral to instruments like spectrometers, interferometers, and refractometers[11]. Traditionally, optical lenses have been made from materials like glass or polymers, relying on their thickness to accumulate optical path differences for phase modulation. However, as technology continues to evolve, the thickness of these lenses has become a limitation, unable to meet the growing demands for integration and miniaturization.

In recent years, the development of micro-nano fabrication technology and new materials has led to the emergence of novel ultra-thin planar lenses. These include several types: metamaterials lenses composed of periodic arrays of metallic wire grids or photonic crystals made from periodically arranged dielectric materials[12] and liquid crystal (LC) lenses based on Pancharatnam-Berry (PB) phase modulation[13]. Liquid crystals are widely used in lens fabrication due to their birefringence and electric controllability. The effective refractive index of liquid crystals for polarized light is determined by the orientation of the molecules, which can be altered by external stimuli such as temperature and electric fields[14,15]. This property provides a significant advantage for liquid crystal in light field modulation. Additionally, while metamaterial lenses are typically manufactured using expensive lithography techniques, the production of LC ultra-thin lenses relies on LC photoalignment technology to control the optical axis direction of LC molecules[13], making it cost-effective and promoting the extensive development of liquid crystal lenses.

For a typical liquid crystal Pancharatnam-Berry (LC PB) lens, the change in the state of polarization (SOP) is induced by the phase retardation between orthogonal polarization components. This phase retardation can be controlled by an external electrical signal and ultimately converted into amplitude modulation through external optical elements, such as polarizers[16]. The handedness of the incident circularly polarized light (CPL) results in opposite phase distributions, leading to a bifocal characteristic[17]. Geometric phase lenses are a special type of lens made using half-wave retardation devices. The optical axis on the lens is not in the same direction but distributed as concentric rings along the radial angle of the surface. Circularly polarized light passing through the wave plate acquires a new phase, with the phase information $\Gamma$ determined by the angle $\alpha$ between the optical axis of the half-wave plate and the lens. This relationship is given by $\Gamma = 2\alpha$. When $\Gamma = \frac{\pi r^2}{\lambda f}$, the condition for forming a liquid crystal geometric phase lens is satisfied, where $r$ is the lens radius, $f$ is the focal length, and $\lambda$ is the design wavelength. Combining these relationships, the following equation can be derived: $\alpha(r) = \frac{\pi r^2}{2\lambda f}$. This formula gives the $\alpha$ distribution on the surface of a lens with focal length $f$ [18]. According to the formula, these lenses act as positive lenses for left circularly polarized light and negative lenses for right circularly polarized light[19]. In recent years, LC PB lenses have seen significant development, particularly in the field of tunable optical phase modulation, drawing considerable attention. These lenses have found applications in various areas, including biomedical imaging[20,21], tunable zoom[22], optical vortices[23], aberration correction[24], and multifocal systems[25,26]. However, research on liquid crystal negative polarization lenses, which are capable of focusing all states of polarization in the same direction, remains limited.

In this paper, we design a liquid crystal hyperbolic lens(LCHL) whose orientation angle can be expressed as $\alpha = \frac{\pi x^2}{2\lambda f_1} - \frac{\pi y^2}{2\lambda f_2}$. It is evident that this lens has different focal lengths in the x and y directions, enabling isotropic focusing for left- and right-handed circular polarization states. We fabricated and studied two lenses with the same focal length and different focal lengths, respectively, and analyzed their diffraction patterns. Our results demonstrate that hyperbolic lenses not only detect the polarization state of light but also have significant advantages in detecting weak polarization states. Therefore, we believe that hyperbolic lenses will become an important tool in future applications of engineering optics and material science.

When a wave plate with an optical axis oriented at an angle θ and a phase delay is $\Gamma$, illuminated by circularly polarized light $E_{in} = \begin{bmatrix} 1 & \sigma i \end{bmatrix}^T$ ($\sigma = \pm 1$ denotes the chirality of circularly polarized light.), the output light's Jones vector can be expressed as $E_{out} = \cos\frac{\Gamma}{2}\begin{bmatrix} 1 \\ \sigma i \end{bmatrix} - i\sin\frac{\Gamma}{2}exp(i2\sigma\alpha)\begin{bmatrix} 1 \\ -\sigma i \end{bmatrix}$. For the case of a half-wave plate, the output light simplifies to a form that $E_{out} = -i\exp(i2\sigma\alpha)\begin{bmatrix} 1 \\ -\sigma i \end{bmatrix}$, indicates the handedness of the input circularly polarized light is reversed, with an additional $phase = \exp(i2\sigma\alpha)$. This additional phase shift can be modulated by the orientation of the wave plate's optical axis $\alpha$. Therefore, by applying different modulations to the optical axis in the x and y directions, the spatial phase can be indirectly modulated as well.

To construct a liquid crystal lens that can focus both right-handed and left-handed circularly polarized light in the same direction, we design the orientation angle which is given by $\alpha(x,y) = \frac{\pi x^2}{2\lambda f_1} - \frac{\pi y^2}{2\lambda f_2}$, where $f_1$ and $f_2$ represent the focal lengths in the x and y directions, respectively. These also define the focal length for right-handed circularly polarized light, and correspondingly, $f_1$ and $f_2$ describe the focal length for left-handed circularly polarized light. Unlike traditional Pancharatnam-Berry (PB) lenses, which exhibit either focusing or defocusing behavior depending on the chirality of the incident circularly polarized light, our designed hyperbolic lens can focus both left-handed and right-handed circularly polarized light.

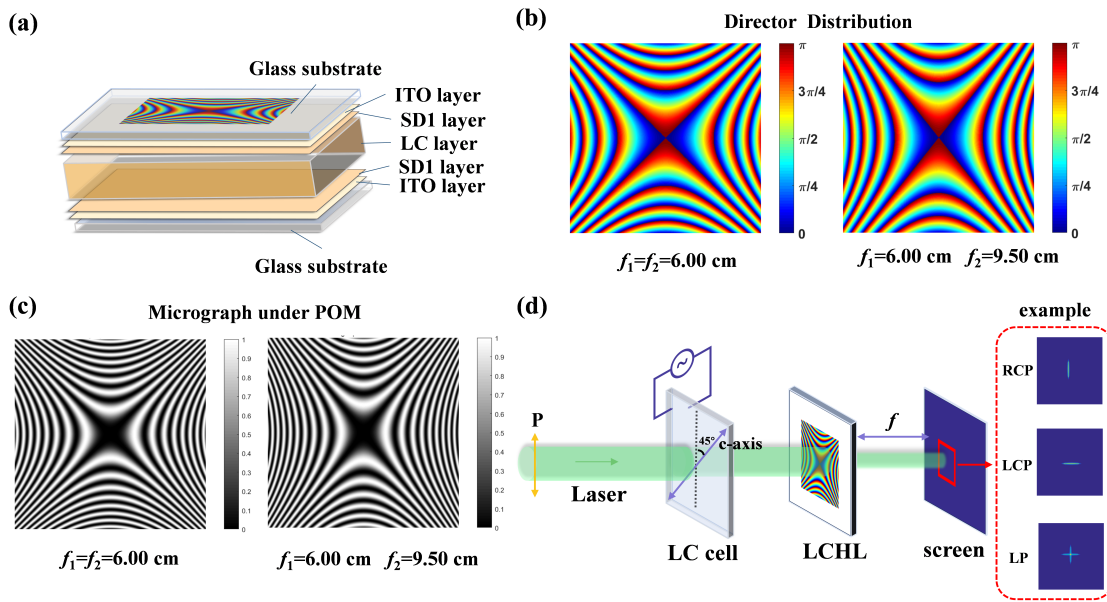

Figure 1.(a)Schematic structure of a hyperbolic lens.(b)Theoretically obtained director distributions for equal and unequal focal lengths.(c)Theoretically obtained micrograph under POM.(d)Schematic diagram of the far-field diffraction optical path

The theoretical design of the lens is shown in Figure 1. (a) illustrates the device structure of the proposed liquid crystal (LC) lens; (b) shows the theoretical director distribution of the liquid crystal molecules for lenses with equal and unequal focal lengths in the x and y directions; (c) presents simulated images of the LC Pancharatnam-Berry (PB) lens observed under crossed polarizers for lenses with equal and unequal focal lengths in the x and y directions.

The optical path setup is shown in Figure 1(d). Linearly polarized light successively illuminates the liquid crystal cell, which has a horizontal director, and a lens sample. The profile of the output beam depends on the angle between the liquid crystal director and the linearly polarized light. When the linearly polarized light is parallel to the liquid crystal director, the light incident on the lens sample remains linearly polarized. At an angle of ±45°, the light becomes right- and left-handed circularly polarized, respectively. The theoretical far-field diffraction patterns at the focal length corresponding to these polarizations are shown in Figure 1(d). In the figure, we have taken the LCHL with $f_1 = f_2$ as an example, the diffraction pattern becomes a vertical line for the right-rotated circularly polarized light and a horizontal line for the left-rotated circularly polarized light, and since the linearly polarized light can be composed of both left-rotated and right-rotated circularly polarized light, it exhibits the pattern of a cross, and the focal length of the lens is determined by the specific value of f.

If $f_1 \neq f_2$, then for $f_1$ it will be the focal length of the right-rotated circularly polarized light, $f_2$ will be the focal length of the left-rotated circularly polarized light, and for the linearly polarized light, the focal length of the diffraction pattern will be the superposition of the diffraction patterns of the left-rotated circularly polarized light and the right-rotated circularly polarized light at that distance.

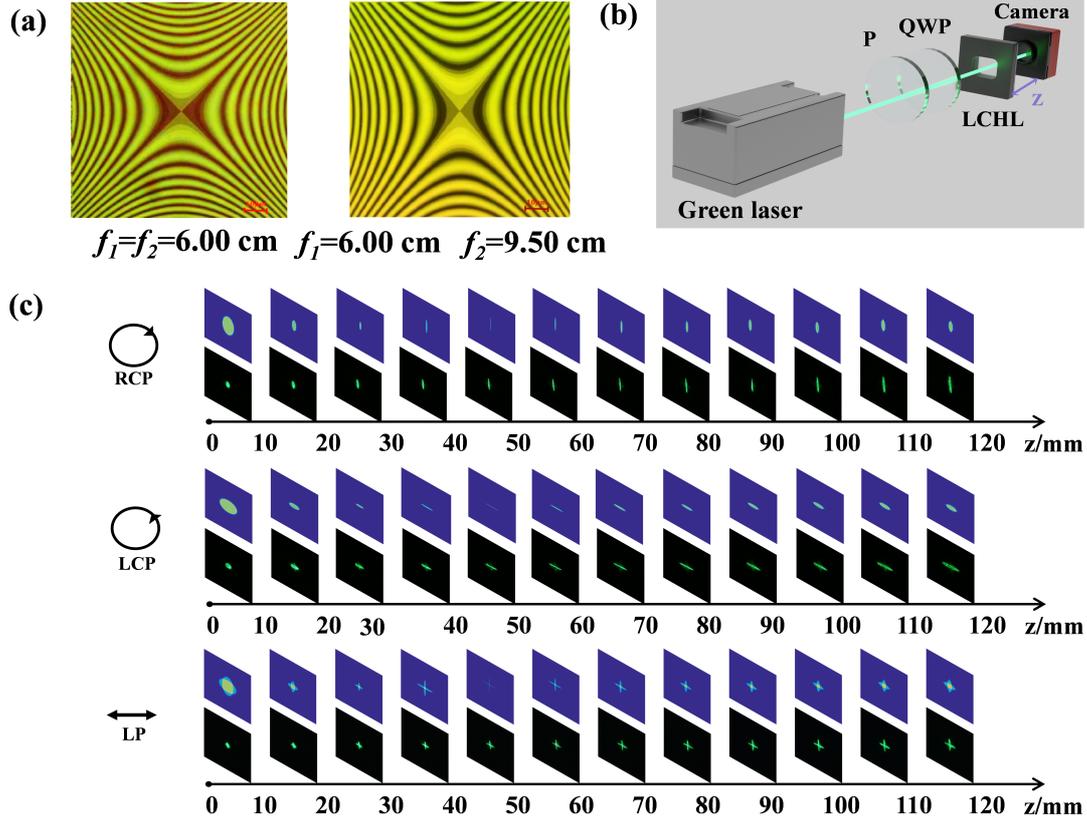

Figure 2.(a)Micrographs of hyperbolic lenses with equal and unequal focal lengths obtained in the experiment under POM.(b)Far-field diffraction optical path design.(d)"Comparison of experimental far-field diffractograms at various focal planes with theoretical predictions for hyperbolic lenses of equal focal lengths under different polarization states

Our lens system imaging test is shown in Figure 2. The optical path of the test setup is illustrated in Figure 2(a).The optical path setup is shown in Figure 2(b), where the polarization state of the incident light is controlled by adjusting the linear polarizer and quarter-wave plate. The light is then directed onto the LCHL, and the far-field diffraction patterns of the output beam are captured by a camera.

In our experimental setup, we used a 532 nm green laser as the light source, with a custom-made LCHL configured such that $f_1 = f_2 = 6.00\ cm$. This configuration allows us to theoretically simulate the far-field diffraction patterns at different focal planes after the incident light passes through the LCHL.

As illustrated in Figure 2(c), we collected diffraction patterns for various focal planes using the $f_1 = f_2 = 6.00\ cm$ lens configuration and compared them with the results obtained from our MATLAB simulations. The comparison shows good agreement between the experimental and simulated results. Specifically, when right-handed circularly polarized light is used, the diffraction pattern forms a vertical line; for left-handed circularly polarized light, the pattern becomes a horizontal line. When linearly polarized light is incident, the diffraction pattern exhibits a cross shape.

In our research findings, for the $f_1 = f_2 = 6.00\ cm$ lens, we determined an experimental focal length of $f = 5.90\ cm$, which closely matches the theoretical focal length.

In summary, we designed and analyzed liquid crystal hyperbolic lenses and demonstrated their unique optical microstructure and properties. This types of lenses are capable of focusing both left- and right-handed circularly polarized light in the same direction, a special focusing behavior achieved through the distribution of the optical axis based on geometric phase principles. The formula that governs the optical axis distribution is key to this functionality, allowing for precise control of the phase and focus. This approach not only applies to our liquid crystal lens design but is also relevant for metasurfaces, offering potential for integration in various advanced optical systems.

Although the experimental investigation of the lens' ability to detect weak polarization states is still ongoing, preliminary results suggest that this hyperbolic lens structure could be highly sensitive to subtle polarization changes, making it a promising tool for weak-light polarization detection in the future.

The experimental results thus far are in good agreement with theoretical simulations, confirming the lens' ability to modulate light fields with high precision. This makes the liquid crystal hyperbolic lens well-suited for applications requiring intricate optical phase modulation, such as in biomedical imaging, optical vortex generation, and multifocal systems. As research progresses, we anticipate that these lenses will play a significant role in the advancement of optical metasurfaces and next-generation optical devices.